\documentclass[conference]{IEEEtran}
\pdfoutput=1

\makeatletter
\usepackage[top= 1 in, bottom=0.75 in, left=0.75 in, right=0.75 in]{geometry}

\usepackage{cite}

\usepackage{graphics,color}
\usepackage{graphicx}
\usepackage{epsfig, psfrag}
\usepackage{subfigure}
\usepackage{epstopdf}
\usepackage{psfrag}
\usepackage[normalem]{ulem}

\usepackage{amsmath}
\usepackage{latexsym}
\usepackage{amssymb}
\usepackage{amsthm} 

\usepackage{placeins}

\usepackage{url}
\usepackage{xspace}
\usepackage{soul}

\newtheorem{theorem}{Theorem}
\newtheorem{lemma}{Lemma}
\newtheorem{proposition}{Proposition}
\newtheorem{corollary}{Corollary}

\newtheorem{fact}{Fact}
\theoremstyle{definition}
\newtheorem{definition} {Definition}

\newtheorem{example}{Example}

\newfont{\boldlarge}{msbm10 scaled 1100}

\newcommand{\ignore}[1]{}

\def\expe{\mathbb{E}}   

\def\P{\mathsf{P}}
\def\ber{\mathsf{Ber}} 
\def\var{\mathsf{Var}} 

\newcommand{\add}[1]{{\color{black}{#1}}}

\newcommand{\kl}[2]{D\left( \left. #1 \right\| #2 \right)}

\newcommand\defeq{\stackrel{\triangle}{=}}

\renewcommand{\qed}{\nobreak \ifvmode \relax \else
      \ifdim\lastskip<1.5em \hskip-\lastskip
      \hskip1.5em plus0em minus0.5em \fi \nobreak
      \vrule height0.2em width0.5em depth0.4em\fi}
\IEEEoverridecommandlockouts
\begin{document}

\sloppy

\title{Reliability of Sequential Hypothesis Testing Can Be Achieved by an 
Almost-Fixed-Length Test}

\author{
 \IEEEauthorblockN{Anusha Lalitha}
  \IEEEauthorblockA{Electrical \& Computer Engineering\\
    University of California, San Diego\\
    Email: alalitha@ucsd.edu} 
\and
  \IEEEauthorblockN{Tara Javidi}
  \IEEEauthorblockA{Electrical \& Computer Engineering\\
    University of California, San Diego\\
    Email: tara@ece.ucsd.edu} 
}

\maketitle

\begin{abstract} 
The maximum type-I and type-II error exponents associated with the newly introduced almost-fixed-length hypothesis testing is characterized. In this class of tests, the decision-maker declares the true hypothesis almost always after collecting a fixed number of samples $n$; however in very rare cases with exponentially small probability the decision maker is allowed to collect another set of samples (no more than polynomial in $n$). This class of hypothesis tests are shown to bridge the gap between the classical hypothesis testing with a fixed sample size and the sequential hypothesis testing, and improve the trade-off between type-I and type-II error exponents. 
\end{abstract}


\section{Introduction}

Statistical hypothesis testing is an integral part of many scientific discoveries and engineering systems. It is also shown to be at the core of 
many problems in information theory and statistics \cite{Csiszar04}. This paper considers the two well-known variants of simple binary hypothesis testing where a decision maker, after observing a sequence of i.i.d random variables, is tasked with identifying the most probable one.  In the first version of the problem the number of samples that is provided to the decision maker is fixed ($\leq n$), while in the second variant of the problem (known as sequential hypothesis testing due to Wald \cite{Wald48}), the decision maker is given the additional freedom to collect a random number of samples so long as the expected number of samples is kept constant ($\leq n$).

There is a large body of literature on the asymptotic analysis of type-I and type-II errors as the (expected) number of samples $n$ grows large. More specifically, the error exponents in both variants of hypothesis testing is well-known and understood \cite{error_exp_blahut, Error_exp_tuncel, hoeffding1948non, Wald48, Chernoff59, CoverBook2nd}. It is well-known that while, in the fixed-length regime, the error exponents of the two types of errors can only be traded-off against each other, the sequential hypothesis tests can achieve both exponents simultaneously. In other words, by allowing the number of samples to be a random number, the sequential hypothesis test resolves the trade-off between  error-types. This suggests that allowing some variability in the number of samples collected is essential for achieving better reliability (error probabilities).  The main contribution  of this paper is to demonstrate that this flexibility need not be significant. More specifically, this paper introduces a new class of hypothesis testing problems, referred to as \emph{almost-fixed-length hypothesis tests} in which the number of samples is kept fixed ($\leq n$) for almost all sample-paths except for an exponentially rare set for which the number of samples collected are allowed to be somewhat larger (bounded by a polynomial function of $n$). We show that this slight flexibility in the sample collection is sufficient to relax the tension and trade-off between the type-I and type-II errors.  

The proposed achievability scheme is a simple two-phase test whose sample-size random variable is equal either  to $n$ or 
$(k+1)n$, where $k$ is an appropriately chosen integer, with the probability of the latter  approaching zero exponentially fast. In other words, the proposed achievability scheme is an almost-fixed-length test whose variance approaches zero as the number of samples collected grows. We also note that the proposed achievability scheme does not require a full sequential computation, and hence, is not as computationally cumbersome as the optimal sequential ratio test. In other words, neither growing variability nor  the computational complexity of sequential ratio tests are essential to obtaining the optimal sequential error exponents.   Our converse proof closely follows a pair of papers by Grigoryan et. al.~\cite{multi_hyp_rej} and Sason~\cite{mod_dev_HT_Sason} on hypothesis testing with rejection, which, in our opinion, have not received their due attention.

\underline{Notation:} Let $\mathbb{R}_+(\mathbb{Z}_+)$, $\mathbb{R}^+(\mathbb{Z}^+)$ denote the non-negative real numbers (integers) and strictly positive real numbers (integers). For a set $S$ and scalar $a \in \mathbb{R}$, $a+S$ denotes the set $\{x + a: x \in S \}$ and  $aS$ denotes the set $\{ax: x \in S \}$. For sets $S_1, S_2$, $S_1 \times S_2$ denotes the set $\{(x_1,x_2):x_1 \in S_1, x_2 \in S_2\}$. Finally, the Kullback--Leibler (KL) divergence between two probability density functions $\P_1(\cdot)$ and $\P_2(\cdot)$ on space $\mathcal{X}$ is defined as $\kl{\P_1}{\P_2}=\sum_{\mathcal{X}} \P_1(x) \log\frac{\P_1(x)}{\P_2(x)}$,
with the convention $0 \log \frac{a}{0}=0$ and $b \log \frac{b}{0}=\infty$ for $a,b\in [0,1]$ with $b\neq 0$.

\section{Problem Formulation}
\label{sec:formulation}

Consider two hypotheses $H_1$ and $H_2$ which correspond to the two possible underlying distributions, $\P_1$ and $\P_2$, governing the samples. In other words, we have
\begin{align*}
H_1 : X \sim \P_1(\cdot),
\quad \text{and} \quad
H_2 : X \sim \P_2(\cdot),
\end{align*}
where $X$ takes values in a finite set $\mathcal{X}$. Consider collecting $\tau$ number of i.i.d samples, where $\tau$ is a random stopping time with respect to the underlying filtration given by $\sigma(X_1, X_2, \ldots, X_n)$. \add{The expectation under hypothesis $H_i$, for $i \in \{ 1, 2\}$, is denoted by $\expe_i[\cdot]$.}

A general \emph{hypothesis test} decides between $H_1$ and $H_2$, for any given $\tau$ samples by dividing the sample space $\mathcal{X}^{\tau}$ into two sets or two ``decision regions''. A decision region, denoted by $A_i^{\tau}$, is a collection of samples $X^{\tau} \in \mathcal{X}^{\tau}$ for which the test chooses $H_i$, for $i \in \{1, 2\}$. The type-I error is defined as an error event that occurs when the test accepts hypothesis $H_2$ when hypothesis $H_1$ is true and its probability is given by $\P_1 \left( A_2^{\tau} \right)$. Similarly, type-II error is defined as an error event when the test accepts hypothesis $H_1$ when hypothesis $H_2$ is true and its probability is given by $\P_2 \left( A_1^{\tau} \right)$. It is known that growing the number of samples results in an exponential reduction in these probabilities of error. This fact is characterized by two classical asymptotic results depending on the manner in which $\tau$ grows.

\underline{Fixed-Length Hypothesis Testing:} In this setting $\tau$ is assumed to be a bounded integer i.e., it satisfies $\tau \leq n$, where $n \in \mathbb{Z}^+$. The error exponents $(E_1, E_2)$ are said to be achievable in a fixed-length setting, if for every $\delta > 0$ there exists a hypothesis test satisfying the following constraints
\begin{align}
&\tau \leq n,
\\
&\P_1 \left( A_2^{\tau} \right)
\leq e^{-(E_1 - \delta)n},
\\
&\P_2 \left( A_1^{\tau} \right)
\leq e^{-(E_2 - \delta)n},
\end{align}
for all $n$ large enough, i.e. $n \geq n_0(\delta)$. 

\begin{definition} 
For any $\lambda \in [0, 1]$, the $\lambda$-tilted distribution $\P_{\lambda}$ with respect to $\P_1(\cdot)$ and $\P_2(\cdot)$ is given by
\begin{align*}
\P^{(\lambda)}(x) 
\defeq
\frac{\P_1^{1-\lambda}(x) \P_2^{\lambda}(x)}{ \sum_{a \in \mathcal{X}} \P_1^{1-\lambda}(a) \P_2^{\lambda}(a)}, \quad \forall \, x \in \mathcal{X}.
\end{align*}
\end{definition}

The following fact characterizes the set of all error exponents, $\mathcal{R}_{FD}$, achievable in fixed-length.

\begin{fact}[Theorem~11.7.1 in \cite{CoverBook2nd}]
The set of error exponents feasible for the class of fixed-length hypothesis tests is  given by 
\begin{align*}
\mathcal{R}_{FD} = 
\left\{ (E_1, E_2): \right.
E_i \leq \kl{\P^{(\lambda)}}{\P_i}, \, i \in \{ 1, 2\},
\nonumber
\\
\hspace{1cm}
\left. \text{ for some } \lambda \in [0,1] \right\}.
\end{align*}
Furthermore, the following fixed-length test achieves the optimal error exponents on the boundary of $\mathcal{R}_{FD}$. If\\
\begin{align*}
\begin{array}{ll}
\frac{1}{n}\sum_{i = 1}^{n} \log \frac{\P_1(X_i)}{\P_2(X_i)} \geq \alpha & \text{ Stop and choose } H_1,
\\
\frac{1}{n}\sum_{i = 1}^{n} \log \frac{\P_1(X_i)}{\P_2(X_i)} < \alpha & \text{ Stop and choose } H_2,
\end{array}
\end{align*}
where $\alpha$ is given by
\begin{align*}
\alpha = \kl{\P^{(\lambda)}}{\P_2} - \kl{\P^{(\lambda)}}{\P_1}, \, \lambda \in [0,1]. 
\end{align*}
\end{fact}

\begin{definition}
Let $\lambda^*$ be such that
\begin{align*}
\kl{\P^{(\lambda^*)}}{\P_1} = \kl{\P^{(\lambda^*)}}{\P_2}.
\end{align*}
Then, the Chernoff exponent $D^*$ is defined as
\begin{align*}
D^* \defeq \kl{\P^{(\lambda^*)}}{\P_1},
\end{align*}
and it characterizes the optimal reliability of Bayesian tests.
\end{definition}

\underline{Sequential Hypothesis Testing:} In this setting, $\tau$ is allowed to be a random variable (potentially unbounded) such that \add{$\max \{ \expe_1[\tau], \expe_2[\tau] \} \leq n$, where $n \in \mathbb{Z}^+$.} The error exponents $E_1$ and $E_2$ are said to be sequentially achievable, if for every $\delta > 0$, there exists a sequential test that satisfies the following
\begin{align}
\label{eq:seq_1}
&\add{ \max \{\expe_1[\tau], \expe_2[\tau] \} \leq n},
\\
\label{eq:seq_2}
&\P_1 \left( A_2^{\tau} \right)
\leq e^{-(E_1 - \delta)n},
\\
\label{eq:seq_3}
&\P_2 \left( A_1^{\tau} \right)
\leq e^{-(E_2 - \delta)n},
\end{align} 
for $n$ large enough, i.e. $n \geq n_0(\delta)$. The following fact characterizes the set of all error exponents, $\mathcal{R}_{seq}$, achievable in sequential manner.

\add{ Our definition of sequentially achievable error exponents, given by equations~(\ref{eq:seq_1})--(\ref{eq:seq_3}),  coincides with the achievable error exponents under~\cite{Csiszar04}. 
Alternatively, the random stopping time $\tau$ can be bounded under each hypothesis as $\expe_1[\tau] \leq n_1$ and $\expe_2[\tau] \leq n_2$, where $n_1, n_2 \in \mathbb{Z}^{+}$ and each error is bounded as $\P_1 \left( A_2^{\tau} \right)
\leq e^{-(E_1 - \delta)n_1}$ and $\P_2 \left( A_1^{\tau} \right)
\leq e^{-(E_2 - \delta)n_2}$, as considered in~\cite{PolyanskiyITA2011}. In contrast, only the case where $n_1 = n_2$ is considered in~\cite{Csiszar04}. The definition in~\cite{Csiszar04} is more stringent than the definition considered in~\cite{PolyanskiyITA2011}. For instance this definition does not admit sequential tests that increase the reliability under $H_1$ by taking arbitrarily large number of samples under $H_1$ than under $H_2$, i.e., by making $\frac{n_1}{n_2}$ arbitrarily large.}

\begin{fact}[Wald and Wolfowitz, \cite{Wald48}]
The set of error exponents feasible for the class of sequential hypothesis test are given by
\begin{align*}
&\mathcal{R}_{Seq} 
\nonumber
\\
= 
&\add{\{E_1: E_1 \leq \kl{\P_2}{\P_1}\} \times \{ E_2:
E_2 \leq \kl{\P_1}{\P_2} \}}.
\end{align*} 
Furthermore, the following sequential hypothesis test achieves the above optimal error exponents $(\kl{\P_2}{\P_1}, \kl{\P_1}{\P_2})$. At any instant $k \in \mathbb{Z^+}$,
\begin{align*}
\begin{array}{ll}
\sum_{i = 1}^{k} \log \frac{\P_1(X_i)}{\P_2(X_i)} \geq  \alpha & \text{Stop and choose } H_1,
\\
\sum_{i = 1}^{k} \log \frac{\P_1(X_i)}{\P_2(X_i)} \leq \beta & \text{Stop and choose } H_2,
\\
\beta < \sum_{i = 1}^{k} \log \frac{\P_1(X_i)}{\P_2(X_i)} < \alpha & \text{ Take an extra sample}\\
 & \text{ and repeat for } k+1,
\end{array}
\end{align*}
where $\alpha = (\kl{\P_2}{\P_1} - \delta)n$ and $\beta = -(\kl{\P_1}{\P_2} - \delta)n$.
\end{fact}

In summary, an optimal fixed-length hypothesis test can only achieve the maximum error exponent in one type of error if the probability of the other error-type is kept fixed. In contrast, a sequential hypothesis test achieves both optimal error exponents simultaneously. Figure~1 illustrates this. 

\begin{example} Consider $H_1: X \sim \ber(0.9)$ and $H_1: X \sim \ber(0.2)$. Figure~\ref{fig:classical_HT} shows the optimal error exponents in both fixed-length and sequential setting. We can see that the sequential hypothesis test provides a significant improvement over the fixed-length hypothesis testing. We shall return to this example to illustrate as how one can go from the fixed-length curve to the sequential curve.

\begin{figure}[!htb]
\label{fig:classical_HT}
\centering
    \includegraphics[width=0.5\textwidth]{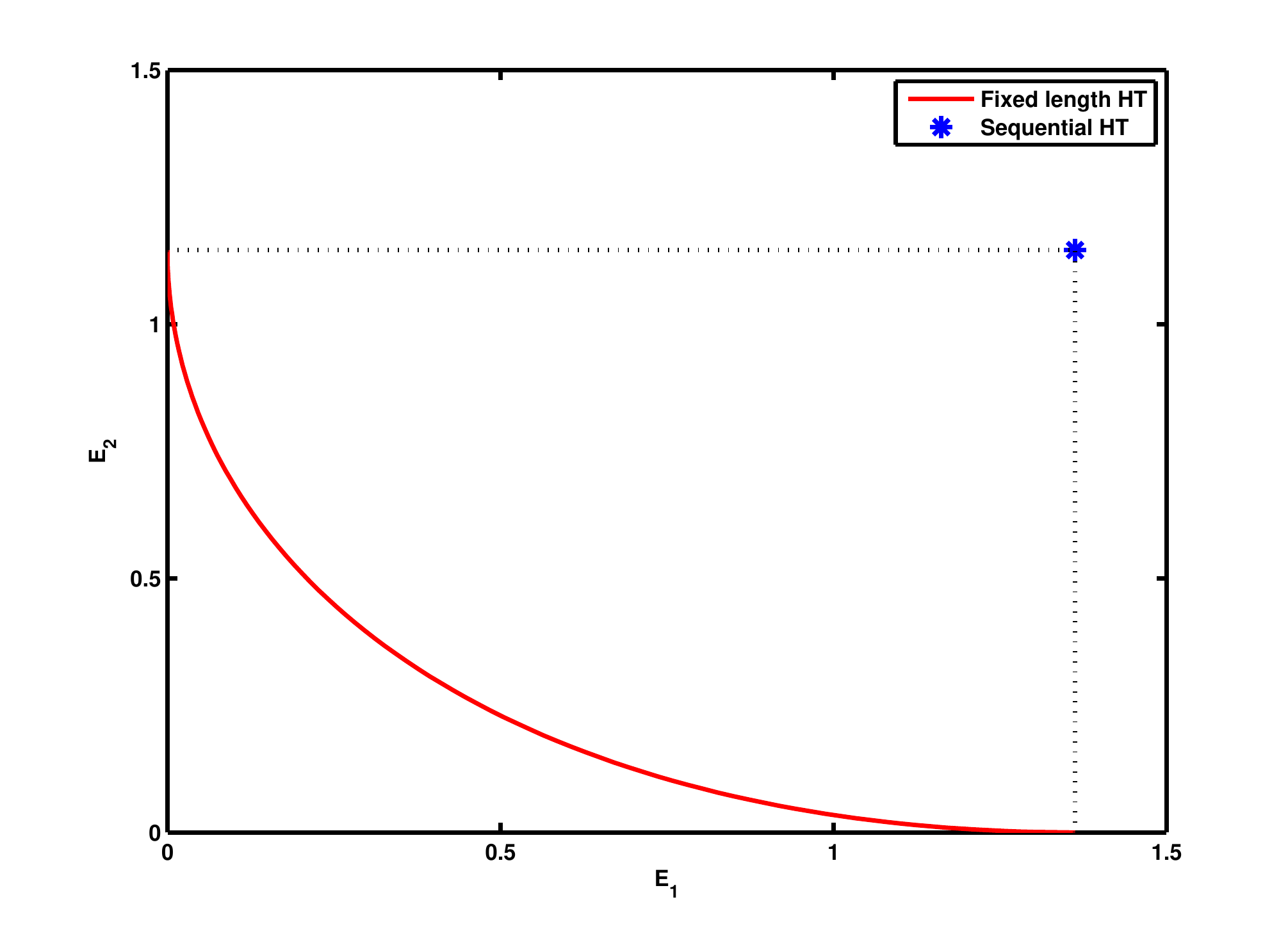}
    \caption{Figure shows the optimal error exponents of fixed-length hypothesis test and sequential hypothesis test for Bernoulli samples with parameters given by $p_1 = 0.9$ under $H_1$ and $p_2 = 0.2$ under $H_2$.}
\end{figure}
\end{example}

\underline{$\gamma$-Almost-Fixed-Length Hypothesis Testing:} We introduce a new class of hypothesis tests for which the number of samples are bounded but have some variability in terms of stopping. By construction, this new class of $\gamma$-almost-fixed-length hypothesis tests are  given an exponentially small flexibility for $\tau$ to be larger than $n$, while keeping the maximum length of any test to be bounded by a polynomial in $n$. The error exponents $(E_1, E_2)$ are said to be achievable in a $\gamma$-almost-fixed-length manner,  $\gamma \in \mathbb{R}^+$, if for every $\delta > 0$ there exists a hypothesis test that satisfies the following
\begin{align}
&\tau \leq O(n^l),
\\
&\P_i(\tau > n) \leq e^{-\gamma n}, \quad
i \in \{ 1, 2\},
\\
&\P_1 \left( A_2^{\tau} \right)
\leq e^{-(E_1 - \delta)n},
\\
&\P_2 \left( A_1^{\tau} \right)
\leq e^{-(E_2 - \delta)n},
\end{align}
for some $l \in \mathbb{Z}^+$ and $n$ large enough. i.e, $n \geq n_0(\delta)$. Let $\mathcal{R}_{\gamma}$ denote the region of all feasible points of the class of $\gamma$-almost-fixed-length tests. We note that as $\gamma \to \infty$, this class of tests recover the class of fixed-length hypothesis tests, hence $\mathcal{R}_{FD} \subset \mathcal{R}_{\gamma}$, for every $\gamma \in \mathbb{R}_+$. Similarly, for all $\epsilon > 0$ and $n$ large enough, we have that $\expe_i[\tau] \leq n + \epsilon$, for $i \in \{1,2\}$. This implies that $\mathcal{R}_{\gamma} \subset \mathcal{R}_{seq}$.

\section{Main Results}

\begin{theorem}
\label{thm:gamma_opt}
For any $\gamma \in \mathbb{R}_+$, the region of all feasible points for the class of $\gamma$-almost-fixed-length 
hypothesis tests, $\mathcal{R}_{\gamma}$,  is such that
\begin{align*}
\mathcal{R}_{FD}
\cup
\add{\left(\{E_1: E_1 \leq  E_1(\gamma)\} \times \{E_2: E_2 \leq E_2(\gamma)\}\right)} \subset \mathcal{R}_{\gamma} ,
\end{align*}
where
\begin{align*}
E_1(\gamma) \defeq \max_{\lambda \in [0,1]} \left\{ \kl{\P^{(\lambda)}}{\P_1} : \kl{\P^{(\lambda)}}{\P_2} \geq \gamma 
\right\},
\end{align*}
and
\begin{align*}
E_2(\gamma) \defeq \max_{\lambda \in [0,1]} \left\{ \kl{\P^{(\lambda)}}{\P_2} : \kl{\P^{(\lambda)}}{\P_1} \geq \gamma 
\right\}.
\end{align*}
Conversely, for every $\gamma \in \mathbb{R}_+$ we have 
\begin{align*}
\mathcal{R}_{\gamma}
\subset
\mathcal{R}_{FD}
\cup
\add{\left(\{E_1: E_1 \leq  E_1(\gamma)\} \times \{E_2: E_2 \leq E_2(\gamma)\}\right)}.
\end{align*}
\end{theorem}

\begin{corollary}
For $\gamma > D^*$, we have $\add{\{E_1: E_1 \leq  E_1(\gamma)\} \times \{E_2: E_2 \leq E_2(\gamma)\}} \subset \mathcal{R}_{FD}$ and hence,
$\mathcal{R}_{\gamma} = \mathcal{R}_{FD}$.
\end{corollary}

\begin{figure}[!htb]
\label{fig:err_exp_region}
\centering
    \includegraphics[width=0.5\textwidth]{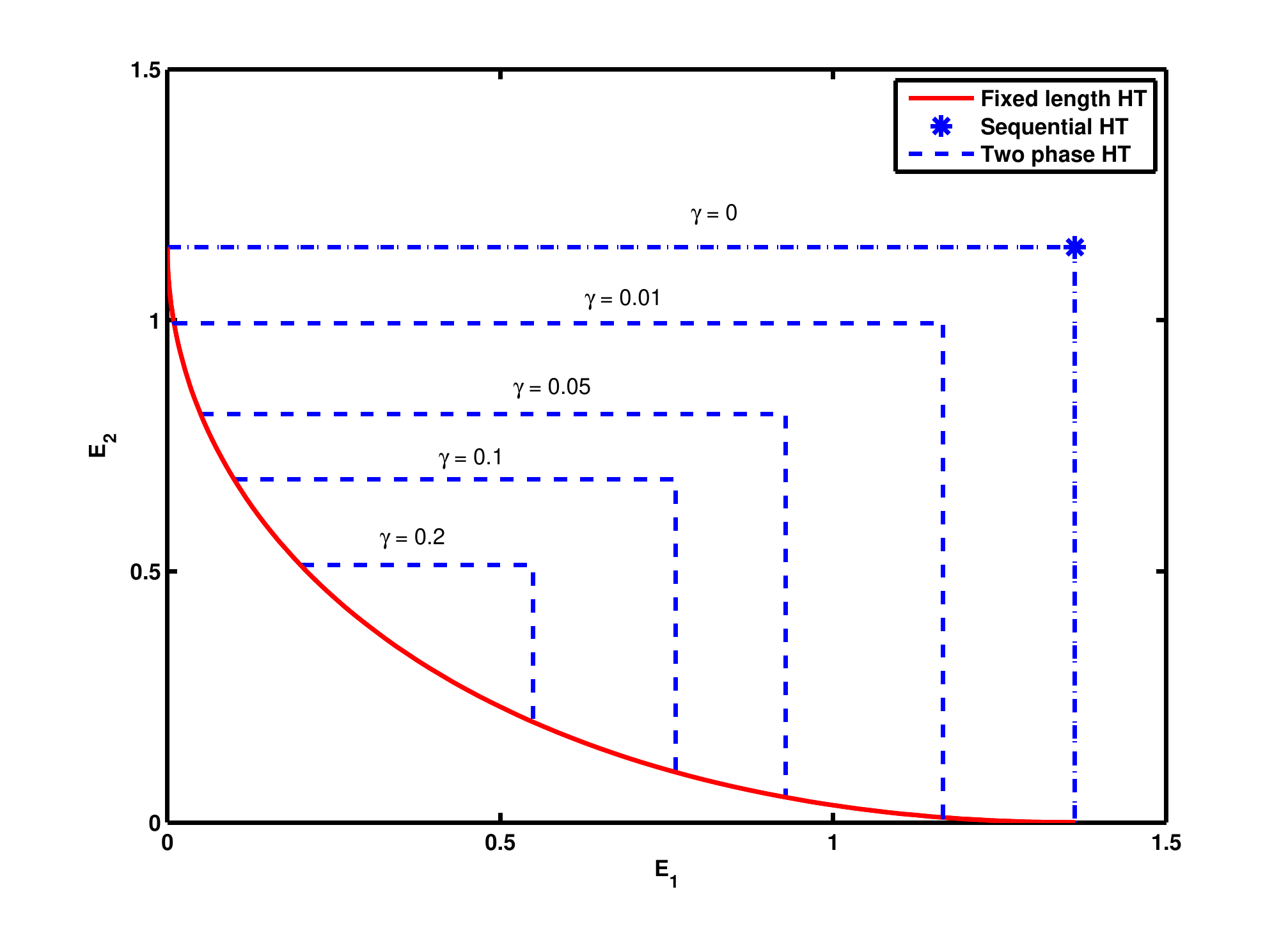}
    \caption{This figure shows the region $\mathcal{R}_{\gamma}$ for various values of $\gamma$ when the samples are Bernoulli with parameters $p_1 = 0.9$ under $H_1$ and $p_2 = 0.2$ under $H_2$. As $\gamma$ decreases the trade-off between the error exponents gets better and the test achieves the optimal sequential exponents $(\kl{\P_2}{\P_1}, \kl{\P_1}{\P_2})$.}
\end{figure}

Figure~2 shows the region of error exponents $\mathcal{R}_{\gamma}$ described in Theorem~\ref{thm:gamma_opt} at different values of $\gamma$. As $\gamma$ decreases, the trade-off between error exponents $(E_1, E_2)$ improves. In particular, it shows that it is possible to achieve the error exponents that are arbitrarily close to optimal error exponents of sequential hypothesis tests, i.e. $(\kl{\P_2}{\P_1}, \kl{\P_1}{\P_2})$, selecting $\gamma$ arbitrarily close to $0$.

\subsection{Achievability: A Two Phase Test}
For $\gamma > D^*$, the achievability of $\mathcal{R}_{\gamma}$ coincides with that of the class of fixed-length hypothesis tests, $\mathcal{R}_{FD}$, ( $\P_i(\tau > n) = 0$ and since $\mathcal{R}_{\gamma} = \mathcal{R}_{FD}$), so any fixed-length hypothesis test achieves $\mathcal{R}_{\gamma}$. 

Let us consider $\gamma\leq D^*$ and $n \in \mathbb{Z}^+$. We propose a hypothesis test that decides between the hypotheses in two phases. In the first phase, we collect $n$ samples and choose whether to stop and decide between the hypotheses or to continue to collect extra  samples. On sample paths where the test continues to the second phase, $k n$ extra samples are obtained, where $k \in \mathbb{Z}^+$ is a fixed parameter of the test. At the end of $(k+1)n$-th instant, the test decides between the hypotheses based on the new $kn$ samples. Hence, this test has two evaluation points, one at $n$ and the other at $(k+1)n$. Formally the two phase hypothesis test is described as follows for $\gamma \leq D^*$.

Let $\lambda_1, \lambda_2 \in [0,1]$  be such that they satisfy the following
\begin{align}
\label{eq:gamma}
\min \left \{ \kl{\P^{(\lambda_1)}}{\P_2},
\kl{\P^{(\lambda_2)}}{\P_1} \right \} = \gamma,
\end{align}
and define $\alpha_1 > \beta_1$ by
\begin{align}
\label{eq:alpha_1}
\alpha_1 = \kl{\P^{(\lambda_1)}}{\P_2} - \kl{\P^{(\lambda_1)}}{\P_1},
\end{align}
\begin{align}
\label{eq:beta_1}
\beta_1 = \kl{\P^{(\lambda_2)}}{\P_2} - \kl{\P^{(\lambda_2)}}{\P_1}.
\end{align}
Note that when $\gamma \leq D^*$, it is always possible to find such $\lambda_1, \lambda_2 \in [0,1]$ and $\alpha_1 > \beta_1$. 

\noindent
\textbf{\underline{Phase I:}} If
\begin{align}
\label{eq:HT_phase1}
\begin{array}{ll}
\frac{1}{n}\sum_{i = 1}^{n} \log \frac{P_1(X_i)}{P_2(X_i)} \geq \alpha_1 & \text{Stop and choose 1},\\
\frac{1}{n}\sum_{i = 1}^{n} \log \frac{P_1(X_i)}{P_2(X_i)} \leq \beta_1 & \text{Stop and choose 2},\\
\beta_1 < \frac{1}{n}\sum_{i = 1}^{n} \log \frac{P_1(X_i)}{P_2(X_i)} < \alpha_1 & \text{Take  extra kn samples}.
\end{array}
\end{align}
Next, define
\begin{align*}
\alpha = \kl{\P^{(\lambda)}}{\P_2} - \kl{\P^{(\lambda)}}{\P_1}, \, \lambda \in [0,1]. 
\end{align*}

\noindent
\textbf{\underline{Phase II:}} If 
\begin{align}
\label{eq:HT_phase2}
\begin{array}{ll}
\frac{1}{k n}\sum_{i = n + 1}^{(k +1)n} \log \frac{P_1(X_i)}{P_2(X_i)} \geq \alpha & \text{Stop and choose 1},\\
\frac{1}{k n}\sum_{i = n + 1}^{(k +1) n} \log \frac{P_1(X_i)}{P_2(X_i)} < \alpha & \text{Stop and choose 2}.
\end{array}
\end{align}

The distribution of random stopping time $\tau$ under hypothesis $H_i$ is given by
\begin{align*}
&\P_i(\tau = l) 
\nonumber
\\
& =
\left\{
\begin{array}{ll}
1 - \P_i\left( \beta_1 < \frac{1}{ n}\sum_{i = 1}^{n} \log \frac{P_1(X_i)}{P_2(X_i)} < \alpha_1 \right) & \text{if } l = n
\\
\P_i\left( \beta_1 < \frac{1}{ n}\sum_{i = 1}^{n} \log \frac{P_1(X_i)}{P_2(X_i)} < \alpha_1 \right) &\hspace{-0.7cm}\text{if } l = kn+n
\\
0  &\hspace{-0.5cm}\text{otherwise}.
\end{array}
\right. 
\end{align*} 
Hence, we have
\begin{align*}
\P_1(\tau > n) 
&=
\P_1\left( \beta_1 < \frac{1}{ n}\sum_{i = 1}^{n} \log \frac{P_1(X_i)}{P_2(X_i)} < \alpha_1 \right) 
\nonumber
\\
&\leq 
\P_1\left( \beta_1 < \frac{1}{ n}\sum_{i = 1}^{n} \log \frac{P_1(X_i)}{P_2(X_i)} \right).
\end{align*}
Using Sanov's Theorem and from equation~(\ref{eq:beta_1}) we have
\begin{align*}
\P_1(\tau > n) 
\leq e^{-\kl{\P^{(\lambda_2)}}{\P_1} n}
 \leq e^{-\gamma n},
\end{align*}
where the last inequality comes from equation~(\ref{eq:gamma}). Similarly, we also have $\P_2(\tau > n) \leq e^{-\gamma n}$. Note that by construction, $\tau \leq (k+1)n$. Hence, this test belongs to the class of $\gamma$-almost-fixed-length hypothesis test.

\begin{proposition}
\label{prop:HT_2phase}
Let $\gamma \leq D^*$ and $k \in \mathbb{Z}^+$. The set of error exponents achieved by the hypothesis test with two phases as given by equations~(\ref{eq:HT_phase1}) and~(\ref{eq:HT_phase2}) is given by
\begin{align*}
\mathcal{R}_{2} = \mathcal{R}_{\gamma} \cap (\gamma + k \mathcal{R}_{FD} ).
\end{align*}

\end{proposition}

Th proof of the above proposition is provided in Appendix-A. 
\begin{corollary}
Define
\begin{align*}
k^* \defeq \max \left \{ \frac{\kl{\P_2}{\P_1}}{D^*}, \frac{ \kl{\P_1}{\P_2}}{D^*} \right \} .
\end{align*}
For all $k \geq k^*$ and $\alpha = 0$, the two phase hypothesis test achieves any $(E_1, E_2) \in \mathcal{R}_{\gamma}$.
\end{corollary}

\subsection{Converse: Hypothesis Testing with Rejection Option}

Our converse bounds the performance of a $\gamma$-almost-fixed-length hypothesis test with that of a fixed-length hypothesis test with rejection option where the probability of rejection approaches zero exponentially fast with an exponent at most $\gamma$. More precisely, a test from the class of hypothesis tests with rejection option, at end of $\tau$ samples divides the sample space $\mathcal{X}^{\tau}$ into three sets or decision regions, given by $A_i^{\tau}$ for $i \in \{1, 2\}$ for which the test accepts $H_i$, and $A_{\Omega}^{\tau}$  which denotes $X^{\tau} \in \mathcal{X}^{\tau}$ for which the test rejects both hypotheses $H_1$ and $H_2$. The exponents $(E_1, E_2, E_{\Omega})$ are said be achievable, if for every $\delta > 0$ there exists a hypothesis test that satisfies the following
\begin{align}
&\tau \leq n,
\\
&\P_1(A_2^{\tau}) \leq e^{-(E_1 -\delta)n}, \quad 
\P_2(A_1^{\tau}) \leq e^{-(E_1 -\delta)n},
\\
&\P_1(A_{\Omega}^{\tau}) + \P_2(A_{\Omega}^{\tau}) \leq e^{-(E_{\Omega} -\delta)n}.
\end{align}

\begin{lemma}
\label{lemma:HT_rej_region}
For any $\gamma \in  \mathbb{R}_+$, let $\bar{\mathcal{R}}_{\gamma}$ denote the region of all feasible error exponents for the class of hypothesis tests with rejection option, then we have
\begin{align*}
\mathcal{R}_{\gamma} \times \{ E_{\Omega}  = \gamma\}
\subset 
\bar{\mathcal{R}}_{\gamma}.
\end{align*}
Conversely, for every $\gamma \in \mathbb{R}_+$ we have
\begin{align*}
\bar{\mathcal{R}}_{\gamma} 
&\subset
\mathcal{R}_{\gamma} \times \{ E_{\Omega}  = \gamma\}.
\end{align*}
\end{lemma}
A variant of above the lemma has been studied under the class of hypothesis tests with rejection option in~\cite{multi_hyp_rej, mod_dev_HT_Sason}.

\begin{corollary}
The following test achieves the optimal error exponents on the boundary of $\bar{\mathcal{R}}_{\gamma}$. If
\begin{align*}
\begin{array}{ll}
\frac{1}{n}\sum_{i = 1}^{n} \log \frac{\P_1(X_i)}{\P_2(X_i)} \geq \alpha & \text{ Stop and choose } H_1,
\\
\frac{1}{n}\sum_{i = 1}^{n} \log \frac{\P_1(X_i)}{\P_2(X_i)} \leq \beta & \text{ Stop and choose } H_2,
\\
\beta < \frac{1}{n}\sum_{i = 1}^{n} \log \frac{\P_1(X_i)}{\P_2(X_i)} < \alpha & \text{ Reject both }H_1, H_2,
\end{array}
\end{align*}
where $\alpha, \beta$ are given by
\begin{align*}
\alpha = \kl{\P^{(\lambda_1)}}{\P_2} - \kl{\P^{(\lambda_1)}}{\P_1},
\end{align*}
\begin{align*}
\beta = \kl{\P^{(\lambda_2)}}{\P_2} - \kl{\P^{(\lambda_2)}}{\P_1},
\end{align*}
where $\lambda_1, \lambda_2 \in [0,1]$ satisfy the following 
\begin{align*}
\min \left \{ \kl{\P^{(\lambda_1)}}{\P_2},
\kl{\P^{(\lambda_2)}}{\P_1} \right \}
 = \gamma.
\end{align*}
\end{corollary}

Note that the first phase of two phase hypothesis test resembles the class of hypothesis tests with a rejection option. In the setting of $\gamma$-almost-fixed-length hypothesis tests, i.e., while $\P_1(\tau > n) \leq e^{-\gamma n}$ and $\P_2(\tau > n) \leq e^{-\gamma n}$, Lemma~\ref{lemma:HT_rej_region} implies that for every $\delta > 0$ and for $n$ large enough, we have $\P_1(A_2^n) \geq e^{- ( E_1 + \delta )n}$ and $\P_2(A_1^n) \geq e^{-(E_2 + \delta)n}$ for $n$ large enough. Since $\P_1(A_2^{\tau}) \geq \P_1(A_2^n)$, for every $\delta > 0$ and for large enough $n$, we have that $-\frac{1}{n}\log\P_1(A_2^{\tau}) \leq E_1 + \delta$. In other words, we have that error exponents of a $\gamma$-almost-fixed-length test are bounded by the error exponents of a hypothesis test with rejection option where $E_{\Omega} = \gamma$. Hence, we have the converse for Theorem~\ref{thm:gamma_opt}.

\section{Conclusion and Future Work}
We looked at a new class of hypothesis tests that have a slight flexibility over fixed-length hypothesis tests by allowing a slightly larger sample size in exponentially small fraction of sample paths. We show that when larger samples are acquired in only exponentially small cases, the overall reliability is increased significantly and the trade-off between type-I and type-II error exponents is relaxed. An interesting area of future work is the optimality of our proposed two-phase scheme when the second phase of the sample collection is limited to $kn$ samples, where $k < k^*$. It is not hard to extend Proposition~\ref{prop:HT_2phase} to arrive at the achievability of this class of tests as shown in Figure~3 (here $k = 2 < 4 = k^*$). However, the converse remains.

\begin{figure}[!htb]
\label{fig:k_2}
\centering
    \includegraphics[width=0.5\textwidth]{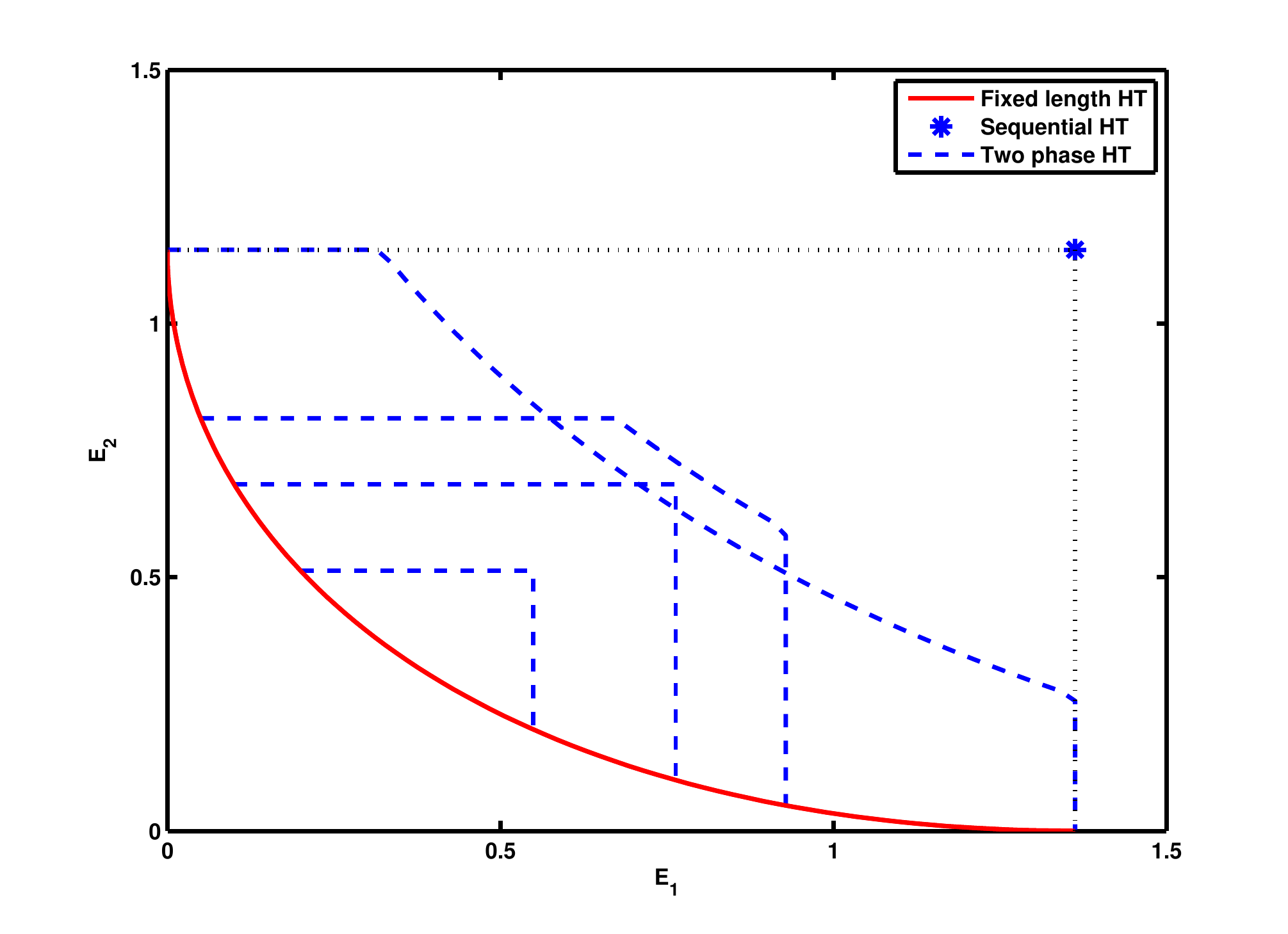}
    \caption{This figure shows the achievable region of the two phase hypothesis test as $\gamma$ increases for k = 2 ($k^* = 4$), when the samples are Bernoulli with parameters $p_1 = 0.9$ under $H_1$ and $p_2 = 0.2$ under $H_2$.}
\end{figure}

Another interesting area of future work is considering the variability of the sample size for various tests. In particular, it is easy to see the following.
\begin{lemma}
\label{lemma:moment}
Let $\tau$ be the number of samples acquired by an $\gamma$-almost-fixed-length hypothesis test where almost all samples are limited to $n$. For $i \in \{1, 2\}$, we have
\begin{align*}
\lim_{n \to \infty}\expe_i \left[\left( \frac{\tau}{n}\right)^l \right] = 1, \, \forall \, l \geq 1, 
\quad
\lim_{n \to \infty} \var_i(\tau) = 0.
\end{align*}
\end{lemma}
\noindent
This means that the class of sequential tests whose variance of stopping time are required to zero is no more restrictive than the class of all sequential tests in terms of reliability.  Similar statements can be made for constraining higher moments of the stopping time. An interesting question is to characterize the optimal error exponents for the class of sequential tests that satisfy the more stringent constraint on the limiting log-moment generating function, i.e.  
\begin{align*}
\frac{1}{\lambda} \log \expe_i[e^{\lambda \tau}] \leq n, \, i \in \{1,2\},
\end{align*}
for some $\lambda \in \mathbb{R}^+$ and $n \in \mathbb{R}^+$. Note that this class of sequential tests ensure that the risk aversion increases with the increase in the average length of the test.

\appendix

%

\subsection{Proof of Proposition~\ref{prop:HT_2phase}}

The error of type-I is given as follows,
\begin{align*}
&\P_1\left( A_2^{\tau}\right)
=
\P_1 \left( \frac{1}{n}\sum_{i = 1}^{n} \log \frac{P_1(X_i)}{P_2(X_i)} \leq \beta_1 \right)
+ 
\nonumber
\\
&
\hspace{0.5cm}\P_1 \left( \beta_1 < \frac{1}{n}\sum_{i = 1}^{n} \log \frac{P_1(x_i)}{P_2(X_i)} < \alpha_1 \right)
\nonumber
\\
&\times
\P_1\left( \frac{1}{k n}\sum_{i = n+1}^{(k+1) n} \log \frac{P_1(X_i)}{P_2(X_i)} < \alpha_2 \right).
\end{align*}
Using Sanov's Theorem and from the definition of $\alpha_1$ and $\beta_1$, for every $\lambda \in [0,1]$, we have
\begin{align*}
&\P_1\left( A_2^{\tau}\right)
\leq
e^{-\kl{\P^{(\lambda_1)}}{\P_1} n} + e^{-\gamma n}e^{-\kl{\P^{(\lambda)}}{\P_1} k n}.
\end{align*}
For every $\delta > 0$, this implies
\begin{align*}
& \frac{1}{n} \log \P_1 \left( A_2^{\tau}\right)
\\
&\leq 
-\frac{\min \{  \kl{\P^{(\lambda_1)}}{\P_1},  \gamma + k \kl{\P^{(\lambda)}}{\P_1}\}}{1 - e^{-\gamma n+ \delta n} + (k+1)e^{-\gamma n} }.
\end{align*}
Now, taking limit we obtain
\begin{align*}
&\lim_{n \to \infty}\frac{1}{n} \log \P_1 \left( A_2^{\tau}\right)
\\
&\leq
- \min \left\{  \kl{\P^{(\lambda_1)}}{\P_1},  \gamma + k \kl{\P^{(\lambda)}}{\P_1}\right\}.
\end{align*}
Similarly, we obtain
\begin{align*}
&\lim_{n \to \infty}\frac{1}{n} \log \P_2 \left( A_1^{\tau}\right)
\\
&\leq
- \min \left\{  \kl{\P^{(\lambda_2)}}{\P_2},  \gamma + k \kl{\P^{(\lambda)}}{\P_2}\right\}.
\end{align*}
Therefore, we have the assertion of the proposition.

\bibliographystyle{IEEEtran}
\bibliography{HypTest}


\end{document}